\documentclass[letterpaper,10pt,english]{IEEEtran}
\usepackage{amsfonts}       
\usepackage{hyperref}
\usepackage{algorithm}
\usepackage{algorithmicx}
\usepackage{amssymb}
\usepackage{amsmath}
\usepackage{graphicx,wrapfig,times,amsthm,bm}
\usepackage{subfig}
\usepackage{threeparttable}
\usepackage[wide]{sidecap}
\usepackage{array}
\usepackage{algorithm}
\usepackage{algpseudocode}
\usepackage{lipsum,babel}
\usepackage{color}
\usepackage{setspace}

\newtheorem{lem}{Lemma}
\newtheorem{deff}{Definition}

\newtheorem{assum}{Assumption}

\def \exp {\mathrm{exp}}

\def \R {\mathbb{R}}

\def \tailcoeff {\beta}

\def \ssconstant {c}

\def \para {\boldsymbol{\theta}}

\def \noise {\xi}

\newcommand{\Mnorm}[2]{{\left\vert\kern-0.30ex\left\vert\kern-0.30ex\left\vert #1 
		\right\vert\kern-0.30ex\right\vert\kern-0.30ex\right\vert}}
\newcommand{\Opnorm}[3]{{\left\vert\kern-0.25ex\left\vert\kern-0.25ex\left\vert #1 
		\right\vert\kern-0.25ex\right\vert\kern-0.25ex\right\vert}_{#2 \to #3}}
\newcommand{\norm}[2]{{\left\vert\kern-0.30ex\left\vert #1 
		\right\vert\kern-0.30ex\right\vert}}
\newcommand{\eigmax}[1]{\left| \boldsymbol{\lambda}_{\max} \left( #1 \right)\right|}
\newcommand{\eigmin}[1]{\left| \boldsymbol{\lambda}_{\min} \left( #1 \right)\right|}

\newcommand{\PP}[1]{\mathbb{P} \left(#1\right)}
\newcommand{\E}[1]{\mathbb{E} \left[#1\right]}

\newcommand{\innerproductminconstant}[1]{\psi_0}

\newcommand{\Kmatrix}[1]{K\left(#1\right)}
\newcommand{\Lmatrix}[1]{L\left(#1\right)}

\newcommand{\estpara}[1]{\widehat{\boldsymbol{\theta}}_{#1}}

\newcommand{\trans}[1]{D_{#1}}
\newcommand{\esttrans}[1]{\widehat{D}_{#1}}

\newcommand{\history}[1]{\mathcal{H}_{#1}}

\newcommand{\SF}[1]{{\bf SF}}
\newcommand{\SP}[1]{{\bf SP}}

\newif\ifarxiv
\arxivtrue

\begin{document}
\ifarxiv
\doublespacing
\onecolumn
\else \fi
\title{Randomized Algorithms for Data-Driven Stabilization of Stochastic Linear Systems}
\author{Mohamad~Kazem~Shirani~Faradonbeh,
	Ambuj~Tewari,
	and~George~Michailidis
	\thanks{}}
\maketitle

\begin{abstract}
	Data-driven control strategies for dynamical systems with unknown parameters are popular in theory and applications. An essential problem is to prevent stochastic linear systems becoming destabilized, due to the uncertainty of the decision-maker about the dynamical parameter. Two randomized algorithms are proposed for this problem, but the performance is not sufficiently investigated. Further, the effect of key parameters of the algorithms such as the magnitude and the frequency of applying the randomizations is not currently available. This work studies the stabilization speed and the failure probability of data-driven procedures. We provide numerical analyses for the performance of two methods: stochastic feedback, and stochastic parameter. The presented results imply that as long as the number of statistically independent randomizations is not too small, fast stabilization is guaranteed. 
\end{abstract}

\section{Introduction} \label{Intro}
Sequential decision-making strategies for stochastic linear systems is an ubiquitous model extensively used in different fields such as biology, robotics, finance, and cryptography \cite{dorato1995linear,burdet2001central,li2004iterative,abeille2016lqg,tsiamis2018information}. The setting consists of a state-space dynamical model following the stochastic trajectory
\begin{eqnarray}
x(t+1) &=& A_0x(t)+B_0u(t)+ \noise(t+1). \label{systemeq1}
\end{eqnarray}
That is, the current state vector $x(t) \in \R^p$ together with the current action $u(t) \in \R^r$ lead the system to the next state $x(t+1)$, according to \eqref{systemeq1}. The linear dynamical model is being disturbed by the stochastic noise process $\left\{ \noise(t) \right\}_{t=1}^\infty$. The state transition matrix $A_0$, and the action influence matrix $B_0$, both of appropriate dimensions, are {\em unknown}.

The objective is to analyze the performance of adaptive stabilization procedures. The focus is on randomized algorithms for stabilizing the system by designing the control actions according to the observed state sequence. Note that the control action can cause instability of the state vector. That is due to the uncertainty of the decision maker about the dynamics matrices $A_0,B_0$. After the stabilization procedure, the control policy for the operating plant under consideration needs to be designed~\cite{guo1991aastrom,ibrahimi2012efficient,faradonbeh2017finite}. The real applications determine the goal being for example minimizing the input energy or regulating the state vector. Therefore, a reliable stabilization algorithm is required to be executed prior to the regulation step, and is desired to guarantee the stabilization quite fast. In general, design of effective methods for finite time stabilization has been recognized as a difficult problem~\cite{li2013stabilization,guo1996global}.

Next, we briefly review various algorithms proposed in the literature for stabilization. Early work consider a restricted setting where the operator norm of the closed-loop matrix is less than one \cite{bittanti2006adaptive}. Then, employing parameter estimation and policy design in an alternate manner, the system is shown to be stabilized~\cite{abbasi2011regret}. Ensuing work establishes the theoretical guarantee for a randomized algorithm based on {\em stochastic feedbacks} (\SF{}) \cite{faradonbeh2018stabilization}. In case of availability of many independent trajectories, stabilizing feedbacks can be learned using only the last sample of each trajectory \cite{dean2017sample}. Further, reinforcement learning algorithms relying on other randomization methods such as {\em stochastic parameters} (\SP{}) \cite{faradonbeh2018optimality}, and input perturbation \cite{faradonbeh2018input,sarkar2018fast} are presented in the literature. The former designs the stabilizing feedback for a randomly generated dynamics parameter, while the latter adds an independent dither signal to the control command. Both procedures then learn the true dynamical parameters through observing the state sequence. 

This work studies the performance of \SF{} and \SP{} which compared to the other aforementioned methods, are able to stabilize a larger class of linear systems. In both algorithms, $k$ random control policies are being applied to the system for episodes of equal lengths.\footnote{$k \geq 1 + \lceil r / p \rceil$ is required, see Section \ref{Algosection} for details} Although some theoretical results exist in the literature~\cite{faradonbeh2018stabilization,faradonbeh2018optimality}, practicality of these stabilization algorithms is not studied. Further, analysis of \SF{} and \SP{} is of independent interest because the underlying stochastic systems are \emph{unstable} and \emph{time-varying}. This work aims to address the above gap through considering the implementation of the algorithms. We provide numerical analysis to compare the amount of time needed for the stabilization. Further, we study the learning accuracy, the failure probability, and the effect of both the number of episodes, and the standard deviation (of the stochastic feedbacks and the stochastic parameters). 

This paper is organized as follows. In Section \ref{model} we formulate the stabilization problem this work is addressing and discuss the theoretical backgrounds. In Section \ref{Algosection}, two randomized algorithms to stabilize the system in finite time are discussed. Then, in Section \ref{simulation} the numerical analysis of the presented algorithms are provided by depicting stabilization speeds, learning accuracies, and failure probabilities.

{\bf Notation:} 
For square matrix $M$, the largest eigenvalue of $M$ (in magnitude) is denoted by $\lambda_{\max}(M)$. For vector $v$, $\norm{v}{2}$ is the Euclidean norm, and the operator norm of matrices is denoted by $\Mnorm{M}{}= \sup\limits_{ \norm{v}{2}=1} \norm{Mv}{2}$. For the notation $\para$, see Definition \ref{notationremark}. Finally, $\mathcal{N}\left(\mu, \Sigma\right)$ denotes the normal distribution with mean vector $\mu$ and covariance matrix $\Sigma$.

\section{Theoretical Framework} \label{model}
To set the stage, we briefly discuss the theoretical framework of the problem. First, the system needs to be stabilizable:
\begin{assum}\label{stabilizability}
	There exists a $r \times p$ stabilizer matrix $L$. That is, $\eigmax{A_0+B_0L} < 1$. 
\end{assum}
The noise in \eqref{systemeq1} is a sequence of centered independent random vectors; $\E {\noise(t)}=0$, satisfying the following. 
\begin{assum}
	The noise covariance matrices are positive definite; $\eigmin{\E {\noise(t)\noise(t)'}}>0$. Further, the noise is sub-Gaussian: $\sup\limits_{t\geq 1} \E{\exp \left({\tailcoeff \norm{\noise(t)}{2}^{2}}\right) }<\infty$, for some $\tailcoeff>0$.
\end{assum}
Given the time length $T$, the objective is to design the control sequence $ \left\{u(t)\right\}_{t=0}^{T-1}$ \emph{effectively}. That is, one can stabilize the system utilizing the observations $ \left\{x(t)\right\}_{t=0}^T$ acquired by applying $ \left\{u(t)\right\}_{t=0}^{T-1}$. For this purpose, we employ a model-based approach so that the algorithm learns $A_0,B_0$ through observing the history by date: $\history{t}=\left\{ \left\{ x(i) \right\}_{i=0}^{t}, \left\{ u(i) \right\}_{i=0}^{t-1} \right\}$. The subsequent argument shows that a coarse approximation suffices for stabilization. To proceed, we define a notation:
\begin{deff} \label{notationremark}
	Henceforth, for all $A \in \R^{p \times p}$, $B \in \R^{p \times r}$, we use $\left[A,B\right]$ and $\para \in \R^{p \times q}$ interchangeably, where $q \equiv p+r$.
\end{deff} 
It is well known that the system can be stabilized by solving a Riccati equation \cite{bertsekas1995dynamic}. First, let $Q \in \R^{p \times p}, R \in \R^{r \times r}$ be arbitrary positive definite matrices. Then, define $\Kmatrix{\para}$ as a possible solution of the algebraic Riccati equation 
$$\Kmatrix{\para} = Q + A'\Kmatrix{\para}A \notag - A' \Kmatrix{\para}B \left(B'\Kmatrix{\para}B+R\right)^{-1} B'\Kmatrix{\para}A.$$
Accordingly, for an arbitrary dynamics parameter $\para$, define the linear feedback matrix 
\begin{eqnarray}
\Lmatrix{\para} &=& -\left(B'\Kmatrix{\para}B+R\right)^{-1} B'\Kmatrix{\para}A. \label{ricatti1}
\end{eqnarray}
The matrix $\Kmatrix{\para}$ exists for almost all $\para$ (w.r.t. Lebesgue measure \cite{abeille2018improved}), and $\Lmatrix{\para}$ stabilizes the system:
\begin{lem}\cite{faradonbeh2018stabilization} \label{optimalityproof}
	There exists a fixed $\epsilon_0>0$, such that an $\epsilon_0$-accurate approximation ensures the stabilization. That is, $\Mnorm{\estpara{}-\para_0}{2} \leq \epsilon_0$ implies that $\eigmax{A_0+B_0\Lmatrix{\estpara{}}} < 1 $.
\end{lem}

\section{Algorithms} \label{Algosection}
The system can be stabilized by leveraging Lemma~\ref{optimalityproof}, together with the recent results on \emph{fast} identification of unstable systems \cite{faradonbeh2018finite,sarkar2018fast}. In the sequel, we explain two different randomization methods, and compare their performance through numerical simulations. Note that since during the stabilization period the system is presumably unstable, the state vector can grow unbounded.

In order to learn the unstable linear dynamical systems, a sufficient and necessary condition is the closed-loop \emph{regularity}~\cite{nielsen2009singular}. In regular matrices, all eigenspaces corresponding to the eigenvalues of magnitude larger than one are \emph{one dimensional}. To satisfy the regularity condition, it suffices to apply stochastic feedbacks~\cite{faradonbeh2018stabilization}. The stabilizing algorithm proposed in the above reference utilizes a few random feedback matrices to accurately learn $\para_0$. 

Namely, suppose that we want to stabilize the system in $T$ time steps. The algorithm applies $k$ statistically independent Gaussian feedback matrices $L_1, \cdots , L_k$ in episodes of the equal length $\lfloor T/k \rfloor$. Collecting the history $\history{k \lfloor T/k \rfloor}$, the algorithm first estimates the closed-loop matrix $A_0+B_0L_i$:
\begin{eqnarray} \label{Closed-Loop-Id}
\esttrans{i} = \arg\min\limits_{\trans{} \in \R^{p \times q}} \sum\limits_{\ell= (i-1)\lfloor T/k \rfloor}^{i \lfloor T/k \rfloor-1} \norm{x(\ell+1)- \trans{} x(\ell) }{}^2.
\end{eqnarray}
Then, the closed-loop estimates $\esttrans{1},\cdots,\esttrans{k}$ are being used to estimate the true dynamics parameter $\para_0$ according to: 
\begin{eqnarray} \label{ParameterLSE}
\estpara{T} = \arg\min\limits_{\para \in \R^{p \times q}} \sum\limits_{i=1}^{k} \Mnorm{ \esttrans{i} - \para \begin{bmatrix} I_p \\ L_i \end{bmatrix}}{}^2.
\end{eqnarray}

\begin{algorithm}
	\caption{\bf: \SF{}} \label{SFalgo}
	\begin{algorithmic}
		\State Input: time length $T$, No. of episodes $k$, variance $\sigma^2$.
		\For{$i=1,\cdots,k$}
		\For{$j=1,\cdots,p$}
		\State Draw column $j$ of $L_i$ from $\mathcal{N}\left(0, \sigma^2 I_r\right)$
		\EndFor
		\For{$ \left(i-1\right) \lfloor \frac{T}{k} \rfloor \leq t < i \lfloor \frac{T}{k} \rfloor$}
		\State Apply $u(t)=L_i x(t)$
		\EndFor
		\State Estimate $\esttrans{i}$ by \eqref{Closed-Loop-Id}
		\EndFor
		\State Compute $\estpara{T}$ according to \eqref{ParameterLSE}
	\end{algorithmic}
\end{algorithm}

The corresponding pseudo-code is provided in Algorithm~\ref{SFalgo}. To ensure that the learning procedure leads to an accurate approximation, the number of episodes $k$ satisfies $k \geq 1+\lceil r/p \rceil$~\cite{faradonbeh2018stabilization}. The following high probability result is established for the performance of the stochastic feedback (\SF{}) method. 
\begin{lem} \cite{faradonbeh2018stabilization} \label{failureprob}
	Defining $\epsilon_0$ similar to Lemma \ref{optimalityproof}, let $\estpara{T}$ be the output of Algorithm \ref{SFalgo}. Then, there is $\ssconstant_0>0$ such that 
	\begin{equation*}
	\PP{ \Mnorm{\estpara{T}-\para_0}{2} > \epsilon_0 } \leq e^{-T \ssconstant_0}.
	\end{equation*}
\end{lem}
Putting Lemma \ref{optimalityproof} and Lemma \ref{failureprob} together, we conclude that the failure probability of \SF{} decays exponentially with $T$.

Here, we also study another randomization procedure based on stochastic parameters (\SP{}). In this method, the algorithm draws Gaussian independent parameters $\para_1, \cdots, \para_k$. Then, in each episode $i$, \SP{} applies the control $u(t)=\Lmatrix{\para_i}x(t)$, which is computed according to \eqref{ricatti1}. At the end of every episode, the closed-loop matrix is estimated by \eqref{Closed-Loop-Id}. Finally, putting the estimates $\esttrans{1}, \cdots, \esttrans{k}$ together, \SP{} learns $\estpara{T}$ by the minimization problem
\begin{eqnarray} \label{SParameterLSE}
\estpara{T} = \arg\min\limits_{\para \in \R^{p \times q}} \sum\limits_{i=1}^{k} \Mnorm{ \esttrans{i} - \para \begin{bmatrix} I_p \\ \Lmatrix{\para_i} \end{bmatrix}}{}^2.
\end{eqnarray}
Similarly, the lower bound $k \geq 1 + \lceil r/p \rceil$ is necessary to ensure identifiability of $\esttrans{T}$. Algorithm \ref{SPalgo} depicts the implementation of \SP{}. 

\begin{algorithm}
	\caption{\bf: \SP{}} \label{SPalgo}
	\begin{algorithmic}
		\State Input: time length $T$, No. of episodes $k$, variance $\sigma^2$.
		\For{$i=1,\cdots,k$}
		\For{$j=1,\cdots,q$}
		\State Draw column $j$ of $\para_i$ from $\mathcal{N}\left(0,\sigma^2 I_p\right)$
		\EndFor
		\For{$ \left(i-1\right) \lfloor \frac{T}{k} \rfloor \leq t < i \lfloor \frac{T}{k} \rfloor$}
		\State Apply $u(t)=\Lmatrix{\para_i} x(t)$
		\EndFor
		\State Estimate $\esttrans{i}$ by \eqref{Closed-Loop-Id}
		\EndFor
		\State Compute $\estpara{T}$ according to \eqref{SParameterLSE}
	\end{algorithmic}
\end{algorithm}

\section{Numerical Analysis} \label{simulation}
In this section we simulate \SF{} and \SP{} for the dynamics matrices
\begin{eqnarray}
A_0 = \begin{bmatrix}
1.07    &     0  & -0.37 \\
0.48  & -0.88 &   0.85 \\
0   & 0.03  & -0.92 
\end{bmatrix}, &&
\:\:\:\:\:\: B_0 =\begin{bmatrix}
-0.48  &  0.44  & -0.29 \\
-0.51   & 0.59  &  0.26 \\
0.29     &    0  & -0.74 \\
\end{bmatrix},  \label{dynamics} 
\end{eqnarray}
and the cost matrices
\begin{eqnarray}
Q = \begin{bmatrix}
1.31  & -0.17  & -0.28 \\
-0.17  &  1.14  &  0.51 \\
-0.28   & 0.51  &  5.01 \\
\end{bmatrix}, &&
\:\:\:\:\:\: R = \begin{bmatrix}
2.01  &  0.54  &  0.77 \\
0.54  &  1.38  &  0.42 \\ 
0.77  &  0.42  &  2.38 \\
\end{bmatrix}. \label{cost}
\end{eqnarray}

Entries of $A_0,B_0,Q,R$ are rounded, after being randomly generated by MATLAB. For this system, solving the Riccati equation, we get
\begin{eqnarray}
\Kmatrix{\para_0} = \begin{bmatrix}
2.83  &  0  & -0.87 \\
0   & 2.20  & -0.32 \\
-0.87  & -0.32  &  7.31 \\
\end{bmatrix}, &&
\:\:\:\:\:\: \Lmatrix{\para_0} =\begin{bmatrix}
0.45  & -0.19 &   0.5 \\
-0.62  &  0.35 &  -0.04 \\
0.13  &  0.06   &-0.77 \\
\end{bmatrix}. \label{RiccSol}
\end{eqnarray}
Then, if one designs the feedback gain according to the true parameter $\para_0$, the spectral radius of the closed-loop matrix is $\eigmax{A_0+B_0\Lmatrix{\para_0}}=0.51$. 

\begin{figure} 
	\centering
	\scalebox{.30}
	{\includegraphics{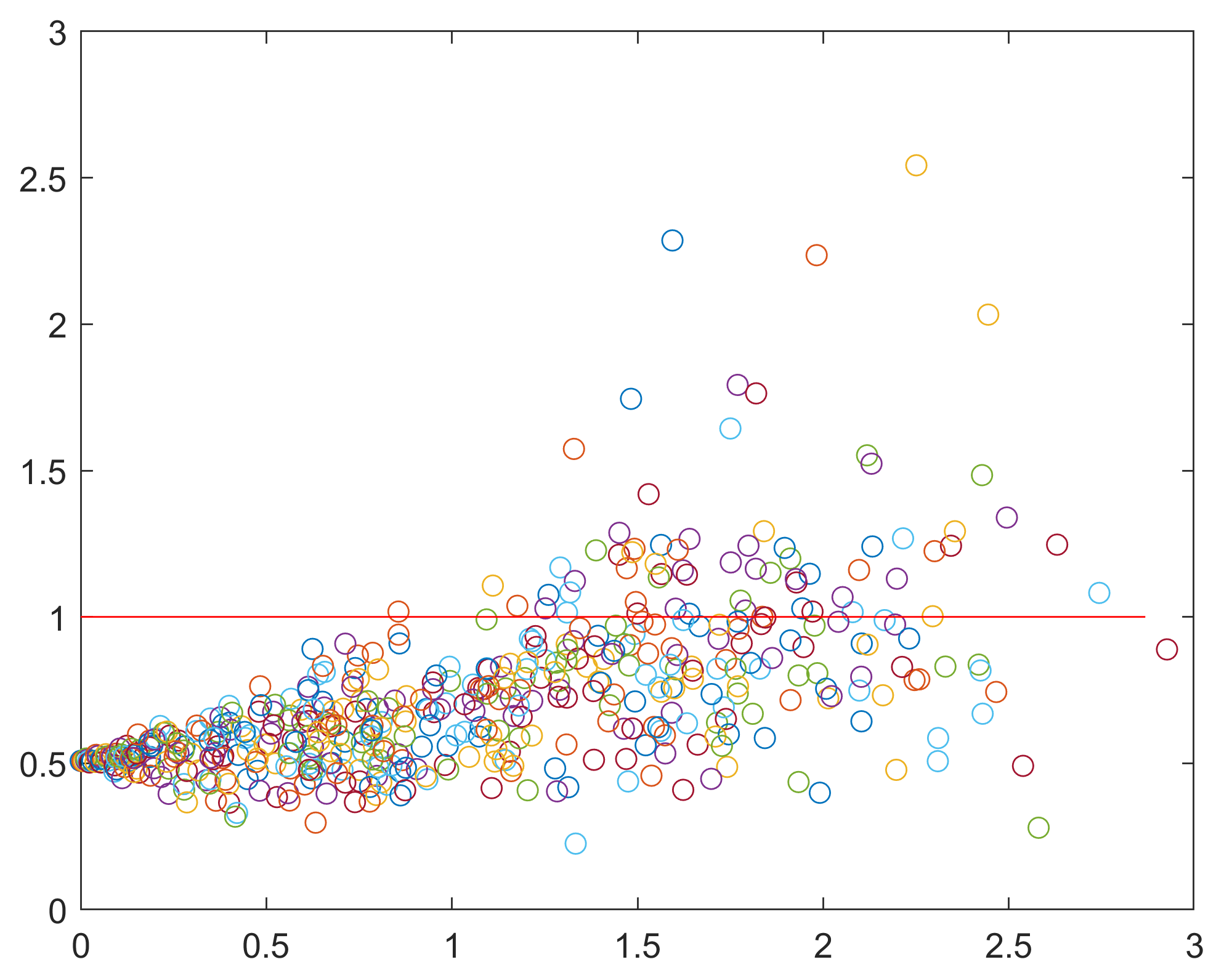}} 
	\caption{$\eigmax{A_0+B_0 \Lmatrix{\estpara{}}}$ vs $\Mnorm{\estpara{}-\para_0}{2}$.}
	\label{Fig1}
\end{figure}
Figure \ref{Fig1} illustrates the statement of Lemma \ref{optimalityproof}. The largest eigenvalue of the closed-loop matrix is reported as a function of the error in the dynamics parameter. The scatter plot consists of different error values $\Mnorm{\estpara{}-\para_0}{2}$, and the resulting spectral radius $\eigmax{A_0+B_0 \Lmatrix{\estpara{}}}$.

\begin{figure}[t!] 
	\centering
	\scalebox{.9}
	{\includegraphics {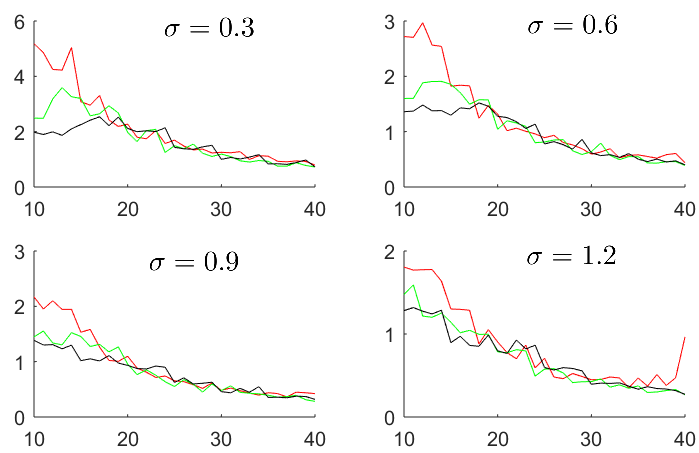}} 
	\caption{$\Mnorm{\estpara{T}-\para_0}{2}$ vs $T$, where $\estpara{T}$ is given by Algorithm \ref{SFalgo}, for $k=3$ (\textcolor{red}{red}), $k=4$ (\textcolor{green}{green}), and $k=5$ (\textcolor{black}{black}).}
	\label{SF-Error}
\end{figure}
\begin{figure}[t!] 
	\centering
	\scalebox{.9}
	{\includegraphics {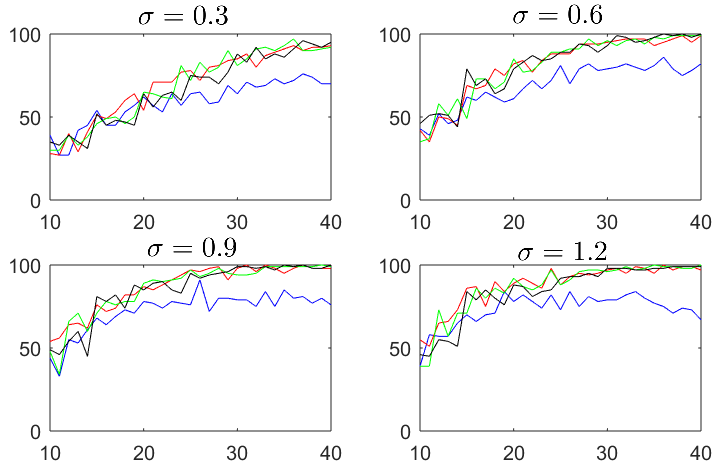}} 
	\caption{Percentage of stabilization by Algorithm \ref{SFalgo} vs $T$, for $k=2$ (\textcolor{blue}{blue}), $k=3$ (\textcolor{red}{red}), $k=4$ (\textcolor{green}{green}), and $k=5$ (\textcolor{black}{black}).}
	\label{SF-prob}
\end{figure}
Figure \ref{SF-Error}, and Fig \ref{SF-prob} plot the performance of Algorithm \ref{SFalgo} for $100$ replications, as the number of time steps $T$ grows. The accuracy of learning the unknown dynamics parameter $\para_0$ is reported in Figure \ref{SF-Error} for different values of $k,\sigma$. Further, Figure \ref{SF-prob} graphs the number of replications in which the system is stabilized. This can be interpreted as a surrogate for the stabilization probability in Lemma \ref{failureprob}. 

\begin{figure}[t!] 
	\centering
	\scalebox{.9}
	{\includegraphics {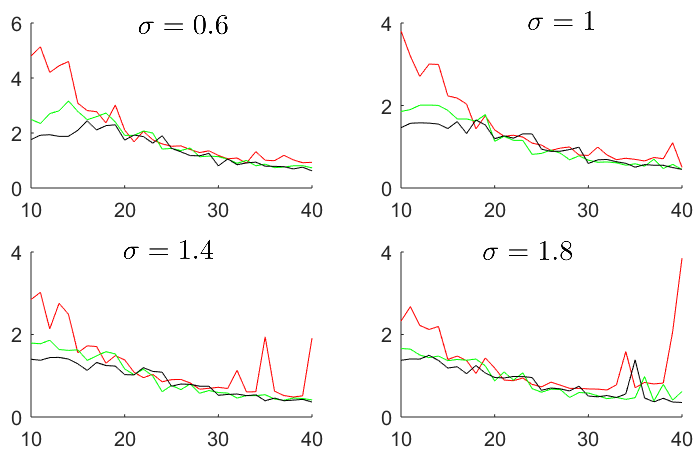}} 
	\caption{$\Mnorm{\estpara{T}-\para_0}{2}$ vs $T$, where $\estpara{T}$ is given by Algorithm \ref{SPalgo}, for $k=3$ (\textcolor{red}{red}), $k=4$ (\textcolor{green}{green}), and $k=5$ (\textcolor{black}{black}).}
	\label{SP-Error}
\end{figure}
\begin{figure}[t!] 
	\centering
	\scalebox{.9}
	{\includegraphics {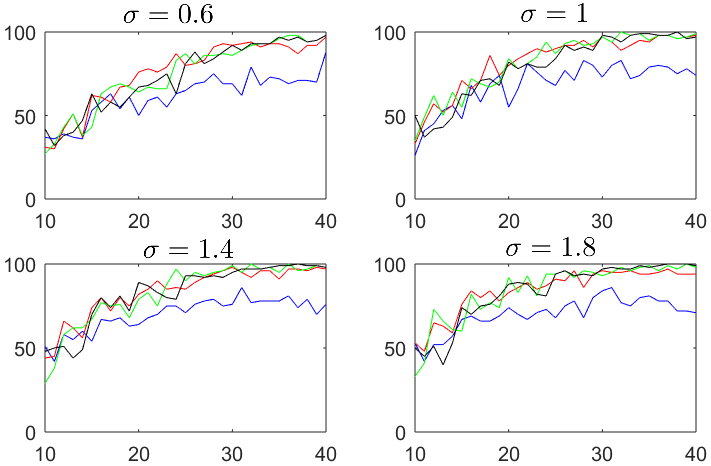}} 
	\caption{Percentage of stabilization by Algorithm \ref{SPalgo} vs $T$, for $k=2$ (\textcolor{blue}{blue}), $k=3$ (\textcolor{red}{red}), $k=4$ (\textcolor{green}{green}), and $k=5$ (\textcolor{black}{black}).}
	\label{SP-Prob}
\end{figure}
Similarly, Figure \ref{SP-Error} and Figure \ref{SP-Prob} illustrate the performance of Algorithm \ref{SPalgo}. The results on the learning error $\Mnorm{\estpara{T}-\para_0}{2}$ are depicted in Figure \ref{SP-Error}. Then, Figure \ref{SP-Prob} plots the percentage of the $100$ replications that the linear feedback designed by the output of Algorithm \ref{SPalgo} stabilizes the true dynamics $\para_0$. Note that in both Figure \ref{SF-Error} and Figure \ref{SP-Error}, the graph for $k=2$ is omitted since in this case the learning error is much larger than the other cases $k=3,4,5$.  

According to the above figures, as long as $k\geq 2$, stabilization will eventually occur. However, the case $k=2$ is significantly slower, and less accurate, compared to the others. All three cases $k=3,4,5$ although, exhibit similar performances. On the other hand, comparing different values of $\sigma$, it is clear that the magnitude of the random matrices ($L_i$ in \SF{} and $\para_i$ in \SP{}) does not have any remarkable influence on the performance. More precisely, different values of $\sigma$ are indifferent as long as they are not too small, or too large. If $\sigma$ is very small, the randomization will be masked by the randomness of the noise process $\left\{w(t)\right\}_{t=1}^\infty$, and is not practical. Further, for large values of $\sigma$, the spectral radius of the resulting closed-loop is wildly large, and leads to an immediate explosion.


\end{document}